\documentclass[conference,twoside]{IEEEtran}
\usepackage{hyperref}
\usepackage{aodv}
\usepackage{url}
\usepackage{breakurl}    
\usepackage{amsfonts,amsmath,amssymb}
\usepackage{algorithm,algorithmic}
\floatname{algorithm}{Process}
\renewcommand{\listalgorithmname}{List of Processes}
\usepackage{graphicx}
\usepackage{color}
\usepackage{pdfsync}
\usepackage[caption=false]{subfig}
\usepackage{ifthen}
\usepackage{xspace}
\newcommand{\tester}{{\small\sf tester}\xspace}
\newcommand{\testers}{{\small\sf testers}\xspace}
\newtheorem{common}{Common}[section]{\bfseries}{\itshape}%
\newcommand{\numberedthing}[2]{%
     \newtheorem{#1}[common]{#2}{\bfseries}{\itshape}}%
\numberedthing{theorem}{Theorem}%
\numberedthing{lemma}{Lemma}%
\numberedthing{prop}{Proposition}%
\numberedthing{remark}{Remark}%
\newtheorem{exam}[common]{Example}%

\newtheorem{definition}[common]{Definition}

\def\squareforqed{\hbox{\rlap{$\sqcap$}$\sqcup$}}
\def\endbox{\ifmmode\squareforqed\else{\unskip\nobreak\hfil
\penalty50\hskip1em\null\nobreak\hfil\squareforqed
\parfillskip=0pt\finalhyphendemerits=0\endgraf}\fi}

\newcommand{\awn}{AWN\xspace}

\makeatletter
\def\comesfrom{\@transition\leftarrowfill}
\def\goesto{\@transition\rightarrowfill}
\def\ngoesto{\@transition\nrightarrowfill}
\def\Goesto{\@transition\Rightarrowfill}
\def\nGoesto{\@transition\nRightarrowfill}
\def\xmapsto{\@transition\mapstofill}
\def\nxmapsto{\@transition\nmapstofill}
\def\@transition#1{\@@transition{#1}}
\newbox\@transbox
\newbox\@arrowbox
\newbox\@downbox
\def\@@transition#1#2%
   {\setbox\@transbox\hbox
      {\vrule height 1.5ex depth .9ex width 0ex\hskip0.25em$\scriptstyle#2$\hskip0.25em}
   \ifdim\wd\@transbox<1.5em
      \setbox\@transbox\hbox to 1.5em{\hfil\box\@transbox\hfil}\fi
   \setbox\@arrowbox\hbox to \wd\@transbox{#1}
   \ht\@arrowbox\z@\dp\@arrowbox\z@
   \setbox\@transbox\hbox{$\mathop{\box\@arrowbox}\limits^{\box\@transbox}$}
   \dp\@transbox\z@\ht\@transbox 10pt
   \mathrel{\box\@transbox}}
\def\nrightarrowfill{$\m@th\mathord-\mkern-6mu%
  \cleaders\hbox{$\mkern-2mu\mathord-\mkern-2mu$}\hfill
  \mkern-6mu\mathord\not\mkern-2mu\mathord\rightarrow$}
\def\Rightarrowfill{$\m@th\mathord=\mkern-6mu%
  \cleaders\hbox{$\mkern-2mu\mathord=\mkern-2mu$}\hfill
  \mkern-6mu\mathord\Rightarrow$}
\def\nRightarrowfill{$\m@th\mathord=\mkern-6mu%
  \cleaders\hbox{$\mkern-2mu\mathord=\mkern-2mu$}\hfill
  \mkern-6mu\mathord\not\mathord\Rightarrow$}
\def\mapstofill{$\m@th\mathord\mapstochar\mathord-\mkern-6mu%
  \cleaders\hbox{$\mkern-2mu\mathord-\mkern-2mu$}\hfill
  \mkern-6mu\mathord\rightarrow$}
\def\nmapstofill{$\m@th\mathord\mapstochar\mathord-\mkern-6mu%
  \cleaders\hbox{$\mkern-2mu\mathord-\mkern-2mu$}\hfill
  \mkern-6mu\mathord\not\mkern-2mu\mathord\rightarrow$}
\makeatother 

\begin{document}%
\title{Modelling and Analysis of AODV in UPPAAL}%

\author{
\IEEEauthorblockN{
Ansgar Fehnker\IEEEauthorrefmark{1}\IEEEauthorrefmark{2},
Rob van Glabbeek\IEEEauthorrefmark{1}\IEEEauthorrefmark{2},
Peter H{\"o}fner\IEEEauthorrefmark{1}\IEEEauthorrefmark{2}, \\
Annabelle McIver\IEEEauthorrefmark{4}\IEEEauthorrefmark{1},
Marius Portmann\IEEEauthorrefmark{1}\IEEEauthorrefmark{3} and
Wee Lum Tan\IEEEauthorrefmark{1}\IEEEauthorrefmark{3}}

\IEEEauthorblockA{\IEEEauthorrefmark{1}NICTA, Australia}
\IEEEauthorblockA{\IEEEauthorrefmark{2}University~of~New~South~Wales, Australia}
\IEEEauthorblockA{\IEEEauthorrefmark{3}The~University~of~Queensland, Australia}
\IEEEauthorblockA{\IEEEauthorrefmark{4}Macquarie University, Australia}
}

\maketitle

\begin{abstract}
This paper describes work in progress towards an
automated formal and rigorous analysis of the Ad hoc On-Demand Distance Vector
(AODV) routing protocol, a
popular protocol used in ad hoc wireless networks.

We give a brief overview of a model of AODV implemented in the  UPPAAL model checker,
and describe experiments carried out to explore AODV's behaviour in two
network topologies. We were able  to locate automatically  and confirm
some known problematic and undesirable behaviours.
We believe this use of model checking as a diagnostic tool complements other formal methods based protocol modelling and verification techniques,
such as process algebras.
Model checking is in particular useful for the discovery of protocol limitations and in the development of improved variations.

\end{abstract}
\hfuzz2pt 

\section{Introduction}
\label{sec:intro}

Route finding and maintenance are critical for the performance of
networked systems, particularly when mobility can lead to highly
dynamic and unpredictable environments; such operating contexts are
typical in wireless mesh networks (WMNs).
Hence correctness and good performance of routing algorithms is highly needed for
routing algorithms. The Ad hoc On-Demand Distance Vector
(AODV) routing protocol~\cite{rfc3561} is a widely used and
adapted routing protocol designed for WMNs and mobile ad hoc networks (MANETs).
It is one of the four protocols
defined in an RFC (Request for Comments) document by the IETF MANET working group.
AODV also forms the basis of new WMN routing protocols, including the upcoming IEEE 802.11s
wireless mesh network standard~\cite{IEEE80211s}.
It has been observed to perform suboptimally~\cite{MK10}, which has provided the impetus to seek variations
and improvements of AODV,
to achieve better overall performance and reliability.
Developing new routing protocols for WMNs and
MANETs, as well as modifications
of existing protocols, is a very challenging task.

Usually, routing protocols are optimised to achieve key
objectives such as providing self-organising capability, overall
reliability and performance in typical network scenarios. Additionally, it is important
to guarantee protocol properties such as loop freedom, for non-typical, unanticipated scenarios. This is particularly relevant for highly dynamic MANETs and WMNs.

The traditional approaches for the analysis of MANET and WMN routing
protocols are simulation and test-bed experiments. While these are
important and valid methods for protocol evaluation, there are
limitations: experimental evaluation is resource intensive,
time-consuming, and even during a very long time of evaluation, only a
finite set of network scenarios can be considered, and no general
guarantee can be given about protocol behaviour for a wide range of
unpredictable deployment scenarios.  The challenges of extensive
experimental evaluation are illustrated by recent discoveries of
limitations of protocols that have been under intense scrutiny over
many years.
A recent example is~\cite{MK10}.
 We believe that formal methods can help in this
regard; they complement simulation and test-bed experiments as methods
for protocol evaluation and verification, and provide stronger and
more general assurances about protocol properties and behaviour.

In this paper we describe work in progress towards the use of model checking
for exploring the behaviour  WMN routing protocols. Model checking is a powerful method, which can be used to validate key correctness properties in finite representations of a formal system model. In the case that a property is found not to hold, the model checker produces evidence for the fault in the form of a  ``counter-example" summarising the circumstances leading to it. Such diagnostic information provides important insights into the cause and correction of these failures.

In \cite{TR11}, we formalised the AODV routing protocol in the process algebra \awn.
In developing the formal specification, we
discovered a number of ambiguities in the IETF RFC \cite{rfc3561}.
Our process algebraic formalisation captures these by several
interpretations, each with slightly different {\awn} code.
Additionally, we found problems with AODV which could lead to undesirable behaviour.
In this paper we employ an interpretation that captures the main intention and
the core functionality of AODV, as defined in~\cite{rfc3561}.
We use the UPPAAL model checker to obtain an executable version of this formalisation.
An executable formal model is  an important tool to confirm and
 discover the presence of and circumstances contributing to  bad behaviour;
it provides insight into network characteristics, e.g. topology forms, which give rise to it.

Deriving an executable model from an existing formal specification is among the
main advantages of our approach.
The {\awn}-specification of AODV is particularly readable, since it  closely
 follows well-known programming constructs. It therefore lends itself
 well for comparison with the original specification of the protocol
 in English.  Based on such a comparison we believe that
 the \awn-model provides a complete and accurate formal
specification of the core functionality of AODV\@.
 By deriving the UPPAAL model from \awn model, this level of trust is
 transferred to the UPPAAL model, which therefore is more reliably a
 correct model of AODV.

The combined use of \awn
 and model checking supports
formal proofs of key correctness
 properties such as loop freedom on the one side, and
formal reasoning with automatic  testing of a large number
topologies and property for a given specification on the other.
Used together, this can provide a powerful tool for the development
and rigorous evaluation of new protocols and variations, and
improvements of existing ones.

The remainder of this paper is organised as follows.
We give a brief overview of AODV
in Section \ref{sec:AodvOverview}.
In Section \ref{uppaal}, we describe the UPPAAL model of AODV, which is
based on a process algebraic model (Section~\ref{ssec:processalgebra}).
Some of our experimental results are presented in Section~\ref{sec:experiments},
and, in Section~\ref{sec:related}, we compare our work to other automated analyses.
Finally, we summarise our work and propose future directions in Section \ref{sec:conclude}.

\section{AODV Overview}
\label{sec:AodvOverview}
AODV is a reactive routing protocol, where the route between the source  and a destination node is established on an on-demand basis. A route discovery process is initiated when a source node $s$ has data to send to a destination node~$d$, but has no valid corresponding routing table entry. In this case, node $s$ broadcasts a route request (RREQ) message in the network. The RREQ message is re-broadcast and forwarded by other intermediate nodes in the network, until it reaches the destination node $d$ (or an intermediate node that has a valid route to node $d$). Every node that receives the RREQ message will create a routing table entry to establish a \emph{reverse route} back to node $s$. In response to the RREQ message, the destination node~$d$ (or an intermediate node that has a valid route to node $d$) unicasts a route reply (RREP) message back along the previously established reverse route. At the end of this route discovery process, an end-to-end route between the source node~$s$ and destination node~$d$ is established. Usually, all nodes on this route have a routing table entry to both the source node~$s$ and destination node $d$. In the event a connection in this end-to-end route were to break down (due to mobility or interference), the node that detects the breakdown sends a route error (RERR) message back to the source node~$s$. The RERR message will cause the set of affected nodes to invalidate their routing table entries that use the broken connection.

\section{Modelling AODV}
Our \ goal is to develop a formal model for
specifying WMN routing protocols in a clear, readable and
unambiguous way. This will allow us to
analyse specifications in a systematic manner, and to compare
variations of particular protocols.

\subsection{Process Algebraic Model of AODV}\label{ssec:processalgebra}
The process algebra \awn~\cite{TR11}
has been developed specifically for modelling WMN routing protocols.\pagebreak{} It is designed in a way to be easily readable and
treats three necessary features of WMNs routing protocols:
 data structures, local broadcast, and prioritised or conditional unicast.
Data structures are used to model routing tables, data packets, etc.;
local broadcast models message sending to {\em all} directly
connected nodes; and the
conditional unicast operator models the
message sending
to one particular node and chooses a continuation process dependent on whether the
message is successfully delivered.

In AWN, broadcast messages are ``guaranteed'', i.e. received by any neighbour that is directly connected.
The abstraction to a guaranteed broadcast enables us to interpret a
failure of  message delivery (under assumptions on the
network topology) as an imperfection in the protocol, rather than as a
result of a chosen formalism not allowing guaranteed delivery. Section \ref{subsec:non-optimal-route}, for example, describes a simple network topology and a scenario for which AODV fails to discover a route. Whenever failure can occur even if broadcast is guaranteed, this means that the failure is a shortcoming of the protocol itself, and cannot be excused by unreliable communication.

Conditional unicast models an abstraction of an ac\-know\-ledg\-ment-of-receipt mechanism
that is typical for unicast communication but absent in broadcast communication, as implemented by
the link layer of relevant wireless standards such as IEEE 802.11. The
process algebra model captures
the bifurcation depending on the success of the unicast, while
  abstracting from all implementation processes of the link layer.

We have used \awn to model AODV according to the IETF RFC3561~\cite{rfc3561}.
The model captures all core functionalities as well as the interface to higher protocol layers via the injection
and delivery of application layer data,  and
the forwarding of data packets at intermediate nodes.
Although the latter is not part of the AODV protocol specification, it is necessary
for a practical model of any reactive routing protocol where protocol activity is triggered via the
sending and forwarding of data packets. In addition, our model contains no ambiguities and no contradictions as they usually
occur in specifications written in natural languages such as in the RFC3561 (see e.g.~\cite{TR11}).

The  model of AODV contains a main process, called {\AODV}, for every node of the network, which handles messages received
and calls the appropriate process {\PKT}, {\RREQ}, {\RREP} or
{\RERR} to handle them. The process also handles the forwarding of any queued data packets
if a valid route to their destination is known. The other processes, like {\RREQ}, handle one particular message type
each. The network as a whole is modelled as a parallel composition of these processes.
Special primitives allow us to express whether two nodes are connected.
Full details of the process algebra description on which our UPPAAL model is based can be found in \cite{TR11}.

\subsection{Modelling AODV in UPPAAL}\label{uppaal}

The process algebraic \awn model of AODV has been used to prove
essential properties, such as loop freedom for a common interpretation
of \cite{rfc3561}. These proofs are valid for
all contexts---independent of a particular topology. However,
during modelling, we discovered a number of possible problems and
unexpected behaviours of AODV. Some required properties for AODV do
not hold, even if guaranteed broadcast is assumed.  For example we
discovered that there is no guarantee that routes will be
established. In this paper, we use model checking to
confirm these problems and to set up an environment to systematically explore and investigate these and other properties. For that reason the UPPAAL model follows faithfully the \awn specification of AODV~\cite{TR11}.

UPPAAL~\cite{uppaal04,LPY97}  is well established and used in particular for protocol verification.
The core of the UPPAAL language consists of networks of timed automata. UPPAAL
provides two synchronisation mechanisms: binary synchronisation channels, which let one automaton synchronise with exactly one other automaton that has an enabled transition with a matching label; and broadcast channels, which let one automaton synchronise with all other automata that have an enabled transition with a matching label.

Broadcast channels are used to model broadcast communication of wireless networks where a message is received by any directly connected node. Every node has a broadcast channel, and every node in range may synchronise on this channel. Connections between nodes are determined by a  ``connectivity" graph, which in UPPAAL is specified by a Boolean-valued function. Every directly connected node  will receive the broadcast message and add it to its message buffer. Broadcast channels are used to model the propagation of {RREQ} and {RERR} messages.

We similarly use UPPAAL's binary (handshake) synchronisation to model
unicast messages. The model includes one channel for each pair of
nodes, and they are only enabled if they are directly connected. If
two nodes are directly connected, nodes are able to receive any
incoming message, and add them to their message buffer. Binary
synchronisation channels are used to model the unicast of  RREP messages and data packets.

Our UPPAAL model is a  direct translation from the model in process algebra to networks of timed automata.
Each node of the network now becomes an automaton\footnote{So far we do not use time and use only finite automata; adding time will be part of future work.} with only one control location (state). This location models the main AODV process. It has local data structures to model the routing table, a buffer for incoming messages to await processing, and a queue for data packets, ready to be forwarded to their respective destinations once routes are found for them. Transitions model the handling of the different types of messages, and thus model the sub-processes {\PKT}, {\RREQ}, {\RREP}, or {\RERR}.
The network as a whole is modelled as a parallel composition of these automata.

Since many problems in AODV arise only under particular circumstances or scenarios,
we also include the possibility of a dynamic topology by allowing connections to be created or deleted. This provides us with the opportunity to explore the more complex behaviours of AODV.

In addition to the timed automata to model individual nodes in the network, we add another process {\tester}, which injects\linebreak\mbox{}

\mbox{}
\vspace*{-36pt}

\noindent
new packets, and manages the connectivity of nodes (create or
delete connections). This additional timed automaton models the environment:
it defines a specific sequence of events and changes to the topology of the network. Using this automaton, we can explore and
analyse particular scenarios. We give examples of \testers below, as well as an illustration of the message sequence charts that are computed by UPPAAL to illustrate violations of a specified property.

With this basic infrastructure, our model consists of one timed automaton per node, plus one node to initiate behaviour.

\section{Experimental Results}
\label{sec:experiments}

In this section we report on two experiments that confirm unexpected and problematic behaviours in the specification of
AODV\@.\footnote{The UPPAAL input can be found at \url{http://www.cse.unsw.edu.au/~ansgar/wripe/}\,.} These problems were discovered during the creation of the  formal
specification in \awn, and we confirm that these problems are also present in the RFC\@.
These ex\-peri\-ments can be considered as validating the UPPAAL model;  in future work we will use
the model checking primarily as a way to explore the types of problems that can arise and under what circumstances.

\subsection{AODV does not always establish routes}\label{subsec:non-optimal-route}

If a node receives a packet from the application layer for an
unknown destination, it initiates a route discovery process. It will
queue the received packet, and broadcast a RREQ message. Nodes that
receive a RREQ message will propagate the message unless either the
node is the destination, or the node has a valid, and fresh enough
routing table entry for the destination. In these cases the node will respond with a RREP message.

In AODV's route discovery process, a RREP message
is unicast back
along a route (usually, established by the RREQ message) towards the
originator of the RREQ message. Every intermediate node on the
selected route will process the RREP message and, in most cases,
forward it towards the originator node. However, there is a
possibility that the RREP message may be discarded at an intermediate
node and result in the originator node not receiving a
reply, and consequently not establishing a route to the destination.

\begin{figure}[h]
 \centering \fbox{
 \begin{minipage}{0.6\linewidth}
 \tester \\[2ex]
\hspace*{0.1\linewidth}\includegraphics[width=0.9\linewidth]{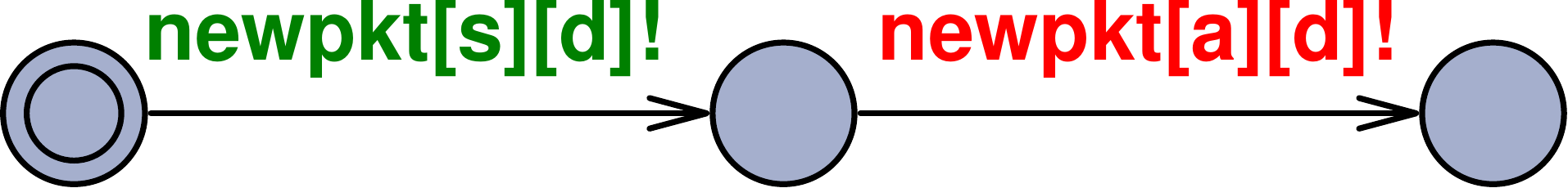}
\end{minipage}
}
\caption{Automaton {\scriptsize\sf tester{}} injecting two new packets.}\label{fig:tester}
\end{figure}

In this experiment we verify that behaviour by setting up a three-node linear topology with nodes $s$, $a$ and $d$.\footnote{
This problem has already been raised on the MANET mailing list in Oct
2004 (\url{http://www.ietf.org/mail-archive/web/manet/current/msg05702.html}).}
Here $s$ is connected to $a$, and $a$ is connected to $d$.
Next we define the UPPAAL process {\tester}, which sets up the following scenario:
 both node $a$ and node $s$ are required to send a data packet to node~$d$ (Fig.~\ref{fig:tester}). Initially all the nodes do not have any information about the connectivity in their routing tables.

The property of interest in this scenario is whether node $s$ is ever able to establish a route to $d$
so that it can deliver its packet. In UPPAAL we use the temporal logic formula
``for all paths there is eventually a (non-empty) routing table
entry at node $s$ for node $d$".
Within UPPAAL's restricted CTL syntax, this property is expressed as:
\\[2mm]\centerline{\tt A<> s.rt[d]\!.nhop!=0}\\[2mm]
The CTL (state) formula {\tt A$\phi$} is satisfied
if all paths starting in that state satisfy $\phi$. The (path)
formula {\tt<>$\phi$} means that $\phi$ holds eventually in some state of a path. The
variable \texttt{s.rt} models the routing table of node $s$, and the field
\texttt{s.rt[d]\!.nhop} represents the next hop for  destination
$d$. If the value of this field is zero, it means that the routing
table entry is empty, or---in terms of the AODV specification \cite{rfc3561}---non-existent.

\begin{figure}[t]
\vspace{4.5pt}
 \centering
\includegraphics[width=0.98\columnwidth]{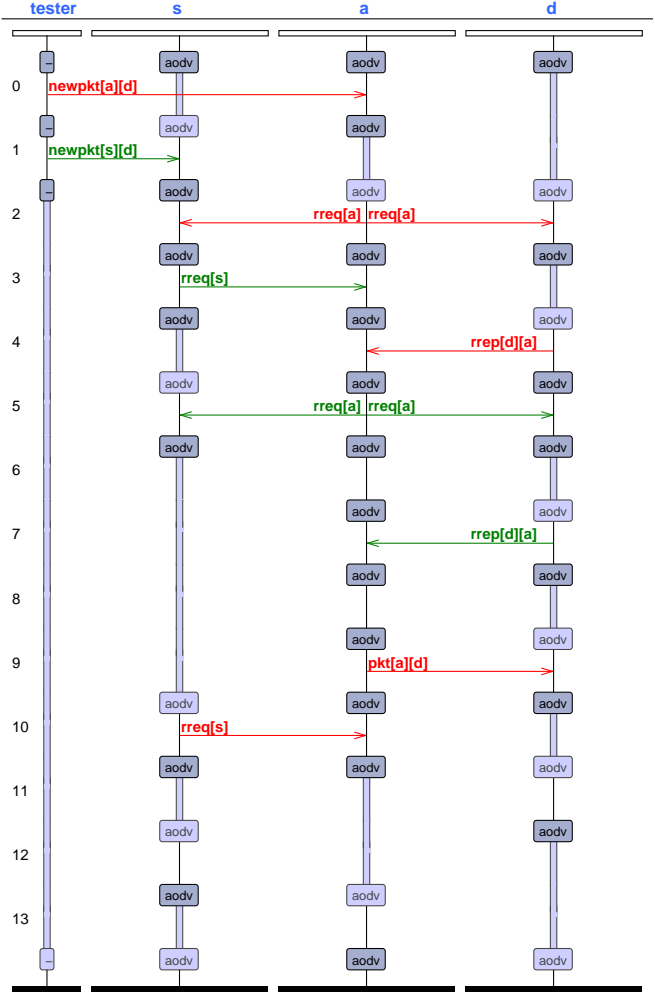}
 \caption{Message Sequence Chart illustrating failed route discovery. Wide vertical lines mean that the local state did not change in this transition for that component.}
 \label{fig:event}
 \vspace{-24pt}
\end{figure}

UPPAAL was able to refute this property, showing that there is indeed
a run where node $s$ is not able to establish a route to node $d$ at all.
The message sequence chart in Fig.~\ref{fig:event} returned by UPPAAL illustrates why this is the case.
It shows the activities and messages sent by the three nodes and the \tester{}.\pagebreak

The scenario begins with two packets both destined for node~$d$, injected by automaton {\tester} at nodes $s$ and $a$ (Steps $0$ and $1$ in Fig.~\ref{fig:event}). Once this has occurred, nodes $s$ and $d$ begin their route discovery process. In this event trace, node~$a$ first broadcasts a route request (Step $2$), which is received by nodes $s$ and $d$; these received messages are placed in the buffers of nodes $s$ and $d$. The message sequence chart only displays the name of the channels that were used, in this case the broadcast channel {\small\textsf{rreq[a]}}. UPPAAL also reports the values of the variables at each node, but it does not include these information in the chart.

Next, node $s$ broadcasts its own route request for $d$, but only $a$ is in range (Step $3$). In the next three events node $d$ responds to $a$'s route request with a route reply message  (Step $4$), then node~$a$ re-broadcasts $s$'s route request (Step $5$), followed by node $d$ responding to node $s$'s route request with a route reply (Step $7$). At this point, AODV (the process running on node~$a$) should forward that reply back to  $s$; however, the remaining events (Steps $8$ to $13$) show that this does not happen.

The problem is that although the request for a route to node~$d$ is properly relayed to the destination $d$, the reply from  $d$ to $s$'s route request is ignored by node $a$. This means that the originator of the second request, node $s$, never receives the reply.  The discarding of the RREP message happens according to the RFC
specification of AODV \cite{rfc3561}. It states that an intermediate node
only forwards the RREP message if it is not the originator node
\emph{and} it has created or updated a route entry to the destination
node described in the RREP message.

\vspace{\abovedisplayskip}
\begin{quote}
\footnotesize{ \sf ``If the current node is not the node indicated by the Originator IP
   Address in the RREP message AND a forward route has been created or
   updated as described above, the node consults its route table entry
   for the originating node to determine the next hop for the RREP
   packet, and then forwards the RREP towards the originator using the
   information in that route table entry.''
   \begin{flushright}
   {[RFC3561, page 21]}
   \end{flushright} }
\end{quote}
\vspace{\belowdisplayskip}

Looking at the sequence chart we see that node $a$ has received a reply to its own request from
node $d$ (Step $4$) and therefore node $a$ has already established a route to $d$ when the
second reply is received in Step $7$. Due to this existing routing table entry, and the fact that the second reply message does not contain any new information about the destination node $d$ (from the point of view of node $a$),
the routing table entry in node $a$ is not updated and therefore the reply is not forwarded. This can be seen
by inspection of the state variables in the UPPAAL model.

A possible solution that would solve this shortcoming would be to forward every reply
received by a node.  Obviously, this would increase the number of control messages generated during route discovery. On the other hand, the AODV specification recovers from this problem by implementing a time-out. If the originator node $s$ does not receive a RREP within a given time, it initiates another route discovery.  The correctness and the efficiency of these two solutions can be compared via simulations. So far, we know  that the time-out solution does not solve the
problem entirely. A repeated route request does not guarantee
the receipt of a route reply and the

\pagebreak
\vspace*{-36pt}

\noindent
property {\tt A<>  s.rt[d]\!.nhop!=0}
still does not hold. It is easy to construct an example;
instead of a linear topology with $3$ nodes, we use a linear topology
with $n+2$ nodes,
where $n$ is the maximum number of repeated route requests~\cite{TR11}.

\subsection{AODV can produce non-optimal routes}
In AODV's route discovery process, a RREQ message is broadcast from a
source node~$s$, in search of a route to the destination node
$d$. Intermediate nodes that receive the RREQ message create routing
table entries for the source node~$s$. In addition, they will
re-broadcast and forward the RREQ message. The destination node $d$ (or an intermediate node that has a valid route to node $d$) will respond with a RREP message and discard the received RREQ message. Duplicate RREQ messages received later (via other routes) are also discarded.  The act of discarding RREQ messages at the destination node can inadvertently cause other intermediate nodes to create non-optimal routes to the source node $s$~\cite{MK10}. Here, we define route optimality in terms of a least-cost metric, i.e., hop count.

In this experiment we confirm that the termination of the route discovery process at the destination node can lead to
other nodes inadvertently creating non-optimal routes to the source node, and that AODV does not always find the best possible routes within the network.

\begin{figure}[ht]
\centerline{
\includegraphics[scale=1.1]{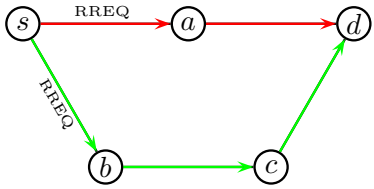}}
\caption{Non-optimal route selection - I}
\label{fig:non-optimal-route}
\end{figure}

We illustrate this phenomenon with the example found by UPPAAL, with
the topology shown in Fig.~\ref{fig:non-optimal-route}. In this
example a new data packet from node $s$ to node $d$ is
injected. Since node $s$ has initially no route to
node $d$, it will initiate the route discovery process. If the
AODV protocol would guarantee an optimal route with respect to hop count, the model should satisfy the following: ``Node $s$ will always find a route to node $d$ with hop count two'', which can be expressed in UPPAAL as
\\[-2mm]\centerline{\tt A<> s.rt[d].hops==2}\\[2mm]
As shown in Fig~\ref{fig:non-optimal-route}, the RREQ messages are forwarded along two paths; via node $a$ and via node $b$.
However, due to some circumstances, node $a$ cannot forward the RREQ message
immediately.  Such situations can occur in real network scenarios due to a number of reasons, such as contention.
In this scenario the RREQ message travelling along the path $s$--$b$--$c$--$d$ reaches
$d$ first. Node $d$ then replies by unicasting a RREP message back to node $s$, via nodes $c$ and $b$.
Only then node $a$ forwards the RREQ message received from node $s$.
Node $d$ discards the RREQ message received from node $a$, since the route request has been handled before.
Hence, node $s$  will establish a non-optimal route (of hop count three) to node $d$, via nodes $b$ and~$c$.

\begin{figure}[ht]
\vspace{3pt}
\centerline{
\includegraphics[scale=1.1]{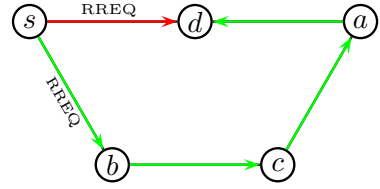}}
\caption{Non-optimal route selection - II}
\label{fig:non-optimal-route2}
\vspace{-10pt}
\end{figure}

Another example of a non-optimal route that does not involve the
timing of RREQ messages was found for the topology shown in
Fig.~\ref{fig:non-optimal-route2}. In this example two new data
packets are injected, first a packet from node $s$ to node $d$,  and
later a packet from node $a$ to $s$.  If the AODV protocol would
guarantee an optimal route with respect to hop count, node $a$ should eventually find a route of hop count two to $s$. The model should guarantee the following:
\\[2mm]\centerline{\tt A<> a.rt[s]\!.hops==2}\\[2mm]
As before, UPPAAL finds a counter-example. First, node $s$ generates and
broadcasts a RREQ message for destination $d$, that is received by its neighbour nodes $b$ and $d$.
Since node $d$ is the destination, it  responds to the RREQ message by  unicasting a RREP message back to $s$, before discarding the RREQ message.
Meanwhile, node $b$ re-broadcasts the RREQ message to its neighbours  $c$ and $s$. Node $c$ then behaves similarly and forwards the
message to $a$ and $b$.  Node $a$ thus establishes a route of hop count three to node $s$ (via $c$ and $b$), before forwarding the RREQ message to $d$. However, node $d$ will drop the RREQ message because it has already received the message previously. If $a$ now has to send a data packet to $s$, it will use the route via $c$ and $b$, rather than the shorter route (of hop count two) via $d$. A possible solution to solve this problem is to allow the destination node $d$ to continue to forward the first RREQ message that it received (as proposed in~\cite{TR11}). By doing so, node $a$ will be able to discover its optimal route (of hop count two) to node $s$, via node $d$.
This proposed solution might have the drawback of increased control message overhead.
However, we do expect the same amount of messages in average---a detailed investigation is part of future work.

The size of the generated UPPAAL models is manageable. The largest of the three models presented in this paper is the last one, with 5 nodes and two injected messages. It has 19833 reachable states. Finding a counterexample to the above specification takes less than 0.1 seconds on an Intel Core i5 2.5 GHz PC.
\section{Related Work}
\label{sec:related}

Other researchers have used formal specification and analysis techniques to investigate the correctness and performance of AODV; we survey the sample  related to model checking.

Bhargavan et al.\ \cite{BOG02} were amongst the first to use model checking on a preliminary draft of AODV, demonstrating the feasibility and value of automated verification of routing protocols. Their investigations using the SPIN model checker revealed that in some circumstances routes containing loops can be created. Their suggested variation which does guarantee loop freedom were not included in the current standard.

Musuvathi et al.\ \cite{MPCED02}  introduced the CMC  model checker
primarily to search for coding errors in implementations of protocols
written in C\@. They use AODV as an example and, as well as discovering
a number of errors, they also found a problem with the specification
itself which has since been corrected.

 Chiyangwa and Kwiatkowska \cite{CK05} use the timing features of
 UPPAAL to study the relationship between the timing parameters and
 the performance of route finding. They were able to establish  a
 dependence between the lifetime of a route and the size of the network, although their study only considered a  single source and single destination, and a simple static linear topology. Our counter-example of Section~\ref{subsec:non-optimal-route} confirms some of the problems that they discovered, and show that they hold even in an untimed model.

 Other researchers have used model checking to analyse other routing protocols. Wibling et al.\ \cite{WPP04} used  SPIN and UPPAAL to verify aspects of the LUNAR protocol, which is also used in ad hoc routing for wireless networks. In particular the timing feature of UPPAAL was used to check upper and lower bounds on route finding and packet delivery times. The scenarios considered included a limited number of topology changes where problems were suspected.

De Renesse and Aghvami \cite{RA04} used SPIN  to study the WARP protocol.  To reduce the overhead on model checking, various simplifications were imposed on a five-node network, including a single source and destination and limitations on the degree that the network can change.

Our approach is in line with these related works.
However, it is unique in the sense that our UPPAAL model complements
our process algebra specification of AODV. As mentioned before, we believe that these two approaches to formal protocol modelling, specification and evaluation, if used together,
can provide a powerful tool for the development and rigorous evaluation of new protocols and variations, and improvements of existing ones.
Currently, our UPPAAL model is derived by hand directly from the \awn process algebra specification, but an automatic translation from \awn in the style of
Musuvathi et al. \cite{MPCED02} would be possible, and remains as future work.

\vspace{1.4pt}
\section{Conclusions and Outlook}\label{sec:conclude}
\vspace{1.4pt}

The aim of our ongoing work is to complement by model checking
a process algebraic description of WMN routing protocols in general, and AODV in particular.
The used description of AODV described in \cite{TR11} is amongst the
first detailed formal models of AODV.
Having the ability of~model~checking
\newpage

formal specifications will allow the confirmation and detailed
diagnostics of suspected errors which arise during modelling.
The availability of an executable model will become especially useful
in the evaluation of proposed improvements to AODV,  something which
we have already started to do \cite{TR11}.

For example, we have sketched a possible solution for the problem
presented in Section~\ref{subsec:non-optimal-route}, which of course
should also be evaluated by formal and rigorous analysis by means of
process algebra and model checking.
We are currently setting up an environment where we can
test a whole bunch of different topologies in a systematic manner. Moreover, we are compiling a library of problem scenarios in AODV. By using the executable model, this will allow us to do a fast comparison between standard AODV and proposed variations in contexts  known to be problematic.
Moreover an executable model will be able to considerably increase the number and variety of scenarios which we can explore.

\bibliographystyle{IEEEtranS}

\end{document}